\documentclass[preprintnumbers,floats,prd,aps,nofootinbib]{revtex4}
\usepackage{amssymb,amsmath}
\usepackage{graphicx,graphics}
\usepackage{subfigure}
\usepackage{hyperref}
\usepackage{float}

\begin{document}
\title{K-Inflation with a Dark Energy Coupling}

\author{Yubei Yue}\email{yueyub@gmail.com}
\author{Bin Chen}\email{bchen01@pku.edu.cn}
\affiliation{
  Department of Physics, and State Key Laboratory of Nuclear Physics and Technology,\\
  Peking University
  Beijing 100081, People¡¯s Republic of China
}
\date{Sep 9, 2009}

\begin{abstract}
It is usually thought that the quintessence as a fundamental
scalar field was already present during the inflationary epoch.
While there are various  models in which the quintessence couples
to other species, it is attractive to anticipate a coupling
between the quintessence and the inflaton in the very early
universe as well. We consider such a coupling in the context of
k-inflation. The coupling function and the potential of the
quintessence are chosen to be of inverse power law forms. We show such a
coupling affects the speed of sound for the inflaton field as well
as the power spectra of perturbations.

\end{abstract}

\maketitle
\section{Introduction} \label{sec:1}

In the context of the inflationary universe scenario \cite{linde
book:1990, Mukhanov book:2005, linde lecture:2006, brandenberger
inflation progress and problems:1999}, the universe experienced a
quasi-exponential expansion at the very early stages of its
evolution, leading to the generation of density perturbations via
quantum fluctuations. These small density perturbations seed the
formation of large scale structure of our universe. Various kinds of
interesting inflationary models could successfully solve the
flatness and homogeneity problems in the hot Big Bang theory.
K-inflation model is one of such candidates. It was first proposed
in \cite{k-inflation:1999, perturbations in k inflation:1999}.
Unlike traditional inflation models with a canonical kinetic term,
in k-inflation paradigm the expansion is driven by a noncanonical
kinetic term. In k-inflation, the potential of the scalar field is not
necessary any more.

In the past decade, several independent but complementary cosmological observations
including supernovae(SN) IA~\cite{SN}, large-scale
structure(LSS)~\cite{LSS}, and cosmic microwave background (CMB)
anisotropy~\cite{CMB}, have provided strong evidences to indicate
that our universe has recently again embarked upon an epoch of
accelerating expansion, driven by so-called dark energy. The
nature of dark energy remains a mystery. The most well-known
candidate of dark energy is the cosmological constant. Even though
such a candidate is still consistent with the current data of
observation, it suffers from fine-tuning and coincidence
problem. Another class of candidates of dark energy is to take the
dark energy as a dynamical scalar field. Among various dynamical dark
energy models\cite{dynamics of dark energy:2006}, quintessence
could be the most famous one\cite{quintessence}. Quintessence is described by an ordinary
scalar field minimally coupled to gravity but with particular potential that
leads to the late time acceleration.

The quintessence field is usually thought to have already existed
during the inflationary stage. Therefore, it is natural to ask
whether there is a coupling between the quintessence field and the
inflaton field, and what is the physical implications of such a coupling.
In \cite{initial conditions:2002}
the authors investigated the initial conditions for the
quintessence field under the assumption that the quintessence
field is sufficiently weakly coupled and is not affected by the
inflaton decay at the end of inflation, they found that the
quintessence field was typically driven to large values. In \cite{large scale cmb and dark energy},
the authors investigated the effects of perturbations in a dark energy
component with a constant equation of state on large scale CMB
anisotropy, and found that the inclusion of the perturbations would
increase the power spectrum. In these studies, there is no direct coupling between dark energy and the
inflation field. Very recently, in
\cite{slow roll inflation with dark energy:2008} the authors
considered a quintessence field coupled directly to the inflaton field with
a square potential. In their model, both the inflation field and the dark energy field were chosen to be canonical scalar fields, and the coupling function was chosen to be exponential form. They found the power spectra of perturbations were slightly modified.

In this paper, we construct a phenomenological model: a k-inflation model with a quintessence
field coupled to it. The coupling function are chosen to be inverse power law form. In our model, the quintessence field has a heavy effective mass, and settles in its effective potential. In next section, we set up our model and
discuss the dynamics of the quintessence field and the evolution of
the scale factor. In section
III, we study the perturbations of our model and calculate the power
spectra. We find that the influence of the quintessence could be
encoded in a modified speed of sound. As an effect, the power
spectra turn to be a little larger. At the end of section
III, we use the WMAP data \cite{five years wmap data:2008} to restrict the parameters in our model. In section IV, we justify
our assumption via numerical analysis. We end with conclusions.

\section{K-Inflation with a Dark Energy Coupling} \label{sec:2}
The action of the theory we consider is
\begin{eqnarray}
{\cal S} = \int {\rm d}^4 x \sqrt{-g}  \left[\frac{M_{\rm
Pl}^2}{2}{\cal R} + {\cal L}_Q + {\cal L}_\phi\right]\nonumber
\end{eqnarray}
with
\begin{eqnarray}
{\cal L}_Q &=& -\frac{1}{2}g^{\mu\nu}(\partial_\mu Q)(\partial_\nu Q) - V(Q) \nonumber \\
{\cal L}_\phi &=& A(Q)F(\phi,X). \nonumber
\end{eqnarray}
Where ${\cal R}$ is the Ricci scalar, and $\phi$ is the inflaton field.
$Q$ is another scalar field which possibly serves the role of dark
energy and $g$ is the determinant of the metric tensor.
$X$ is defined by
\begin{equation} X=-\frac{1}{2}g^{\mu\nu}(\partial_\mu
\phi)(\partial_\nu \phi). \nonumber
\end{equation}
In k-inflation models, the
pressure of the inflaton field is usually written as
$F(\phi,X)$. In this model, we investigate the possibility that the dark
energy field contributes an extra factor $A(Q)$ to the pressure $F(\phi,X)$. In this case, the pressure
of the inflaton field is written as $A(Q)F(\phi,X)$, where Q denotes
the dark energy field. The
equations of motion for the field $\phi$ and Q in a homogeneous
and isotropic universe are
\begin{equation}
(F_X+2XF_{XX})\ddot \phi +3HF_X \dot \phi +(2X\frac{\partial F_X}{\partial \phi}-\frac{\partial F}{\partial \phi}) =-\frac{F_X}{A(Q)}\frac{\partial A(Q)}{\partial Q}\dot Q \dot\phi
,
\end{equation}
\begin{equation}\label{eqQ}
\ddot Q +3H\dot Q +\frac{\partial V(Q)}{\partial Q}-\frac{\partial
A(Q)}{\partial Q}F(\phi,X)=0.
\end{equation}
Here $F_X$ and $F_{XX}$ refer to $\frac {\partial F(\phi,X)}{\partial X}$ and $\frac
{\partial^2F(\phi,X)}{\partial X^2}$ respectively. H is the Hubble parameter,
given by $\frac {\dot a}{a}$.

In this paper, we choose the factor $A(Q)$
to be a dimensionless coupling function
\begin{equation}\label{A}
A(Q)=A_0 (\frac {Q}{M_{\rm Pl}})^{\beta},
\end{equation} where $A_0$ is a constant.
We choose the
potential of the quintessence field to be an inverse power-law
potential,
\begin{equation}\label{V}
V(Q) = V_0 M_{\rm Pl}^4(\frac{Q}{M_{\rm Pl}})^{-\lambda},
\end{equation}
where $\lambda$ is positive, and $V_0$ is a constant. We use the
Planck units $M_{\rm Pl}= 1$ throughout this paper.

The energy density and the pressure of the inflaton field $\phi$ are
\begin{equation}\label{epsilon and p}
\rho = A(Q)(2XF_X-F),  \qquad p=A(Q)F.
\end{equation}
The energy density and the pressure of the Q field are
\begin{equation}
\rho_Q =\frac{1}{2}\dot Q^2+V(Q),  \qquad p_Q=\frac{1}{2}\dot
Q^2-V(Q).
\end{equation}
The Friedmann equations read
\begin{eqnarray}
H^2 &=& \frac{1}{3M_{\rm Pl}^2} (\frac{1}{2} \dot Q^2 +
V(Q)+A(Q)(2XF_X-F)),\label{friedmann1} \\
\dot H &=&  -\frac{1}{2M_{\rm Pl}^2} (\dot
Q^2+A(Q)2XF_X).\label{friedmann2}
\end{eqnarray}

During inflation, the inflaton field dominates the evolution, and the Q field moves in an effective potential. In this
paper we consider the case in which the effective potential for the
Q field has a minimum, and it exists for a positive
$\beta$. The existence of such a minimum was already explored in \cite{slow roll inflation with dark energy:2008}, and the method used here is similar to that in \cite{slow roll inflation with dark energy:2008}. 

From eqn.(\ref {eqQ}), we require
\begin{equation}\label{1Dveff0}
\frac{\partial V_{eff }(Q)}{\partial Q}=\frac{\partial V(Q)}{\partial Q}-\frac{\partial
A(Q)}{\partial Q}F(\phi,X)=\frac{1}{Q}(\beta
A(Q)\overline{F(\phi,X)}-\lambda V(Q))=0,
\end{equation}
where $\overline{F(\phi,X)}=-F(\phi,X)$(during inflation $F(\phi,X)<0$). We have
used expressions for the potential $A(Q)$ and $V(Q)$. This
gives a relation between the potential of the Q field and the
pressure of the inflaton field,
\begin{equation}\label{relation1}
V(Q_{min})=\frac{\beta}{\lambda}A(Q_{min})
\overline{F(\phi,X)}=-\frac{\beta}{\lambda}p.
\end{equation}
From the above equation we can see the value of $\frac{\beta}{\lambda}$ determines the ratio of the energy density of the dark energy field to the inflaton field (during inflation $p+\rho\approx0$). If we require the inflaton field dominates the evolution, the value of $\frac{\beta}{\lambda}$ must be small.
The value of the Q field near the minimum of
the effective potential is
\begin{equation}\label{Qmini}
Q_{min}=\big(\frac{\lambda V_0}{\beta A_0
\overline{F(\phi,X)}}\big)^{\frac{1}{\beta + \lambda} }.
\end{equation}
The expression for $A(Q_{min})$ is
\begin{equation}\label{relation2}
A(Q_{min})=A_0\big(\frac{\lambda V_0}{\beta A_0
\overline{F(\phi,X)}}\big)^{\frac{\beta}{\beta +\lambda}}.
\end{equation}
Eqn.(\ref{friedmann1}) simplifies to
be
\begin{equation}\label{fried1}
H^2  \approx \frac{1}{3}(-\frac{\beta}{\lambda}A(Q)F+A(Q)(2XF_X-F)) \approx
\frac{1}{3}(1+\frac{\beta}{\lambda})A(Q)
(2XF_X-F)\approx \frac{1}{3}(1+\frac{\beta}{\lambda})\rho,
\end{equation}
in the first step we use the eqn.(\ref{relation1}) to eliminate $V(Q)$, and in the second step we use the slow roll condition
\begin{equation}\label{the first flow parameter}
\epsilon\equiv -\frac{\dot H}{H^2}\approx\frac{3}{2}(1+\frac{\beta}{\lambda})^{-1}\frac{2XF_X}{2XF_X-F},
\end{equation}
to add a term $\frac{\beta}{\lambda}A(Q)2XF_X$ to the expression. 

Using eq.(\ref{1Dveff0}), the mass square of the Q field near the minimum of the effective
potential is
\begin{eqnarray}\label{meff}
m^2_{eff}&=&\frac{\partial}{\partial Q}(\frac{1}{Q}(\beta
A(Q)\overline{F(\phi,X)}-\lambda V(Q))) \nonumber \\ &=&
-\frac{1}{Q^2}(\beta A(Q)\overline{F(\phi,X)}-\lambda
V(Q))+\frac{1}{Q^2}(\beta^2 A(Q)\overline{F(\phi,X)}+\lambda^2
V(Q)) \\\nonumber &\approx& \frac{1}{Q^2}(\beta^2 A(Q)(2XF_X-F)+\lambda^2
V(Q)) \\\nonumber &\approx& \frac{\partial^2 V(Q)}{\partial Q^2}+\frac{3H^2}{Q^2}\frac{\beta^2}{1+\frac{\beta}{\lambda}}.
\end{eqnarray}
In the second line, the first term vanishes because of eq.(\ref{1Dveff0}). In the third line, we replace $-A(Q)F$ with $A(Q)(2XF_X-F)$, because of $\rho+p=0$ in the slow roll limit. In the last line we have used eqn.(\ref{fried1}). For
$\beta\ll1$, $\lambda\sim1$ and $\beta^2/Q^2>1$, $m^2_{eff}/3H^2$
exceeds order unit. The range of $Q$ is restricted by experimental data, and is very small in our model. In \cite{stochastic quintessence} the authors give a lower limit of the energy density:
\begin{equation}
\rho_Q \geq 10^{-109}M^4_{\rm Pl}.
\end{equation}   
This corresponds to $Q\leq 1.8\times 10^{-6}M_{\rm Pl}$ with the $\lambda$ and $V_0$ used in section.(\ref{sec:4}).
So even for a $\beta$ as small as $10^{-3}$, the effective mass is rather large, and the $Q$ field would quickly settle in the minimum of its effective potential. This was also justified in
section.(\ref{sec:5}) via numerical analysis.

From eqn.(\ref{Qmini}) we get
\begin{equation}\label{Q velocity}
\dot Q=-\frac{Q}{\beta+\lambda}\frac{\dot F}{F}.
\end{equation}

\section{Perturbations} \label{sec:3}
It is useful to calculate perturbations to see how
the coupling with a dark energy field influences the power spectra
of curvature and tensor perturbations of the inflaton field. Below
we follow the standard procedures in
\cite{perturbations:1992,perturbations in k inflation:1999} and
restrict ourselves in a flat background. First we rewrite eqn.(\ref{friedmann1}) and eqn.(\ref{friedmann2})
\begin{eqnarray}
H^2 &=& \frac{1}{3} (-\frac{\beta}{\lambda}p+\rho)\label{f1} \\
\dot H &=&  -\frac{1}{2} (p+\rho)\label{f2}
\end{eqnarray}
in here, together with a redundant
equation which can be derived using the above two
\begin{equation}\label{f3}
\dot \rho=-3H(p+\rho)+\frac{\beta}{\lambda}\dot p.
\end{equation}
In deriving the above expressions we have neglected the term $\dot Q^2$, and used eqn.(\ref{epsilon and p}) and eqn.(\ref{relation1}). The last expression differs from usual form by an extra term
$\frac{\beta}{\lambda}\dot p$. In this section we adopt the signs and
conventions as in \cite{perturbations in k inflation:1999}. The
energy momentum tensor is
\begin{equation}
T_{\mu\nu}=(\rho+p)u_{\mu}u_{\nu}-(1+\frac{\beta}{\lambda})p
g_{\mu\nu},
\end{equation}
where
\begin{eqnarray}
 u_{\mu}=\frac{\phi_{,\mu}}{(2X)^{\frac{1}{2}}}. \nonumber
\end{eqnarray}
The inflaton field is separated into two parts, one homogenous part
which only depends on time and the other fluctuating part,
\begin{equation}
\phi(t,x)=\phi_0(t)+\delta\phi(t,x).
\end{equation}
The metric is
\begin{equation}
ds^2=(1+2\Phi)dt^2-(1-2\Phi)a^2(t)\gamma_{ik}dx^idx^k.
\end{equation}
The time derivative of the pressure can be obtained by using
eqn.(\ref{f3}) and $\dot\rho=\rho_{,X}\dot
X+\rho_{,\phi}\dot\phi$,
\begin{eqnarray}
\dot p&=&p_{,X}\dot X+p_{,\phi}\dot\phi \\ \nonumber &=&
-3H\frac{c^2_s}{1-c^2_s\frac{\beta}{\lambda}}(\rho+p)+\dot
\phi (\frac{1}{1-c^2_s\frac{\beta}{\lambda}}p_{,\phi}
-\frac{c^2_s}{1-c^2_s\frac{\beta}{\lambda}}\rho_{,\phi}),
\end{eqnarray}
where $c^2_s$ is defined as in \cite{perturbations in k
inflation:1999},
\begin{equation}
c^2_s\equiv
\frac{p_{,X}}{\rho_{,X}}=\frac{\rho+p}{2X\rho_{,X}}.
\end{equation}
The perturbations of the energy and pressure of the inflaton field can
be expressed as
\begin{eqnarray}
\delta\rho=\rho_{,X}\delta
X+\rho_{,\phi}\delta\phi=\frac{\rho+p}{c^2_s}((\frac{\delta\phi}{\dot\phi})\dot\quad-\Phi)
-3H(\rho+p)
\frac{\delta\phi}{\dot\phi}+\frac{\beta}{\lambda}\dot
p\frac{\delta\phi}{\dot\phi},\nonumber
\end{eqnarray}
\begin{eqnarray}
\delta p=p_{,X}\delta
X+p_{,\phi}\delta\phi=(\rho+p)((\frac{\delta\phi}{\dot\phi})\dot\quad-\Phi)+\dot
p\frac{\delta\phi}{\dot\phi}.\nonumber
\end{eqnarray}
We skip from here the index $0$ for the background value of
$\phi(t)$. The perturbations of the components of the energy
momentum tensor can then be expressed as
\begin{eqnarray}\label{cs substitution}
\delta T^0_0=\delta\rho-\frac{\beta}{\lambda}\delta
p=\frac{1-c^2_s\frac{\beta}{\lambda}}{c^2_s}(\rho+p)((\frac{\delta\phi}{\dot\phi})\dot\quad-\Phi)
-3H(\rho+p)\frac{\delta\phi}{\dot\phi},
\end{eqnarray}
\begin{eqnarray}
\delta T^0_i=(\rho+p)(\frac{\delta\phi}{\dot\phi})_i.
\end{eqnarray}
The above two expressions should be substituted into the $00$ and
$0i$ linerized Einstein equations. It is convenient to change from
the independent variables $\Phi$ and $\frac{\delta\phi}{\dot\phi}$
to the new variables $\xi$ and $\zeta$(defined in eqn.(19) and eqn.(20) in \cite{perturbations in k inflation:1999}). The action can be inferred
from the equations for the variables $\xi$ and $\zeta$.

We find if we make the substitution
\begin{eqnarray}\label{cssquare}
\frac{c^2_s}{1-c^2_s\frac{\beta}{\lambda}}\quad\rightarrow\quad \widetilde{c_s}^{2}
\end{eqnarray}
in eqn.(\ref{cs substitution}), the following derivation can be repeated as in the standard case. The symbol $\widetilde{c_s}$ denotes a modified speed of sound.
Further introducing the Mukhanov variable $v=z\zeta$, where $z$ is defined
as
\begin{eqnarray}
z\equiv \frac{a(\rho+p)^{1/2}}{\widetilde{c_s}H}\nonumber
\end{eqnarray}
in a flat universe, we obtain an equation for the variable $v$
\begin{eqnarray}
v^{\prime\prime}-\widetilde{c_s}^{2}\Delta
v-\frac{z^{\prime\prime}}{z}v=0.\nonumber
\end{eqnarray}
In the above equation, the prime represents derivative with respect to the conformal time. The speed of sound is modified in this model, and it recovers the
normal form in the limit of $\beta\rightarrow 0$. We consider
parameters to satisfy $\frac{\beta}{\lambda}\ll 1$, therefore, in this
model we stay in the scope of positive speed of sound. Before we write down the expressions for the power spectra, let's first introduce another two slow roll parameters~(see \cite{Cosmological constraints on general single field inflation}),
\begin{eqnarray}
\eta\equiv \frac{\dot\epsilon}{H\epsilon},  \quad\quad\quad   \kappa\equiv\ -\frac{\dot{\widetilde{c_s}^{-1}}}{H\widetilde{c_s}^{-1}}.
\end{eqnarray}

The power spectrum of curvature
perturbations is
\begin{eqnarray}\label{curvature perturbations}
{\cal P}_{\cal R}\approx \frac{1}{8\pi^2M^2_{\rm
Pl}}\frac{H^2}{\widetilde{c_s}\epsilon} \mid_{aH=\widetilde{c_s}k}.
\end{eqnarray}

The scalar spectral index is
\begin{eqnarray}
n_S-1=\frac{d\ln {\cal P}_{\cal R}}{d\ln k}\approx
-(2\epsilon+\eta+\kappa).
\end{eqnarray}

Tensor power spectrum
\begin{eqnarray}
{\cal P}_{g}\approx \frac{2H^2}{\pi^2M^2_{\rm
Pl}}\mid_{aH=k}.
\end{eqnarray}

The tensor to scalar ratio is
\begin{eqnarray}
r\equiv\frac{{\cal P}_{g}}{{\cal P}_{\cal
R}}=16\widetilde{c_s}\epsilon.
\end{eqnarray}

The tensor spectral index is
\begin{eqnarray}
n_T=\frac{d\ln {\cal P}_{g}}{d\ln k}\approx
-2\epsilon.
\end{eqnarray}

These expressions are similar to the standard case, but both $\widetilde{c_s}$ and $\epsilon$ in these expressions are slightly modified, also the Hubble parameter $H$ contains an extra factor $A(Q)$.

\section{Constraining the parameters}\label{sec:4}
In this section we set up to estimate the parameters in our model. First we need to specify an expression for $F(\phi,X)$. Several forms of $F$ have been already used in the literature. The form $F(\phi,X)=f(\phi)(-X+X^2)$ was first discussed in \cite{k-inflation:1999}, where an inverse square pole-like $f(\phi)$ led to a power law k-inflation. A discuss about such form can also be found in \cite{Self-reproduction in k-inflation:2006}, where the authors argued that for a general form of $f(\phi)=\phi^s$ with $s>0$, inflation could be successfully produced. Here we choose
\begin{eqnarray}\label{the function of F}
F(\phi,X)=f(\phi)(-X+X^2),\nonumber  \quad\quad\quad  f(\phi)=\phi^2.
\end{eqnarray}
In slow roll limit, $X$ is fixed in the point $X_0=\frac{1}{2}$ to satisfy $F_X=0$. In post slow roll stage, we can expand $X$ to $X_0+\delta X$.

The model has four parameters, namely $\lambda$, $\beta$, $A_0$, and $V_0$. First one can use eqn.(\ref{Qmini}) to eliminate Q in eqn.(\ref{fried1}) to get
\begin{eqnarray}
H^2\approx\frac{1}{3}(1+\frac{\beta}{\lambda})A_0\big(\frac{\lambda V_0}{\beta A_0\overline{F}}\big)^{\frac{\beta}{\beta +\lambda}}(2XF_X-F)\approx\frac{1}{3}(1+\frac{\beta}{\lambda})A_0\big(\frac{\lambda V_0}{\beta A_0}\big)^{\frac{\beta}{\beta +\lambda}}\frac{1}{4}f(\phi).
\end{eqnarray}
For convenience, we set
\begin{eqnarray}
C=\frac{1}{3}(1+\frac{\beta}{\lambda})A_0\big(\frac{\lambda V_0}{\beta A_0}\big)^{\frac{\beta}{\beta +\lambda}}.
\end{eqnarray}
And $H^2$ writes as
\begin{eqnarray}\label{hi}
H^2=\frac{C}{4}f(\phi).
\end{eqnarray}
The requirement of $60$ e-foldings of expansion is
\begin{equation}\label{N}
N=\int H dt=\int \frac{H}{\dot\phi}d\phi=\frac{\sqrt C}{4}(f(\phi_i)-f(\phi_e))=60.
\end{equation}
Here and below the subscripts $i$ and $e$ denote the value at the beginning and the end of inflation.
Using eqn.(\ref{f3}), one can get the derivative of $X$ in term of $\phi$,
\begin{equation}
(\dot X)_i=\frac{1}{4\phi_i}.
\end{equation}
Expanding eqn.(\ref{f3}) to the first order in $\delta X$, and using
\begin{equation}
F_X= F_{XX} \delta X=2f(\phi)\delta X,
\end{equation}
one gets
\begin{equation}\label{deltaX}
\delta X=\frac{1}{6\sqrt C f(\phi)}.
\end{equation}
The inflation ends when $\frac{\delta X}{X_0}=1$,
and one gets
\begin{equation}
f(\phi_e)=\frac{1}{3\sqrt C}.
\end{equation}
Using eqn.(\ref{N}), the value of $f(\phi_i)$ is
\begin{equation}\label{phii}
f(\phi_i)=\frac{1}{\sqrt C}(4N+\frac{1}{3}).
\end{equation}
By substituting it back into eqn.(\ref{deltaX}), one has
\begin{equation}
\delta X_i=\frac{1}{24N+2}.
\end{equation}
We can express the value of speed of sound and the slow roll parameters at the beginning of inflation in term of $\delta X_i$ and $\dot X$,
\begin{equation}
(c_s^2)_i=\frac{F_X}{F_X+2XF_{XX}}=\delta X_i, \quad \epsilon=12(1+\frac{\beta}{\lambda})^{-1}\delta X_i, \quad \eta=\frac{\dot X_i}{H_i\delta X_i}, \quad \kappa=\frac{\dot X_i}{2H_i\delta X_i}.
\end{equation}
All the above parameters are evaluated at the beginning of inflation.
The WMAP data of the amplitude of curvature perturbations $\Delta^2_{\cal R}$ is $2.41\times10^{-9}$, and we have
\begin{equation}\label{amplitude of curvature perturbations}
8\pi^2M^2_{\rm
Pl}\widetilde{c_s}\epsilon\Delta^2_{\cal R}=H^2.
\end{equation}
One gets
\begin{equation}
H_i\approx6.46\times10^{-6}.
\end{equation}
By substituting eqn.(\ref{hi}) into eqn.(\ref{phii}), one gets
\begin{equation}
C=\frac{H_i^4}{(N+\frac{1}{12})^2}=4.82\times10^{-25}, \quad \phi_i=(\frac{4H_i^2}{C})^{\frac{1}{2}}=7.32\times10^{9} .
\end{equation}
One can also get $\phi_e=6.93\times10^{5}$ and $H_e=2.40\times10^{-7}$. In \cite{Exploring Parameter Constraints on Quintessential Dark Energy: the Inverse Power law model} and references there, flat inverse power law model of dark energy is favored by the current data, for example $0.5\leq\lambda\leq 2$. As is discussed below eqn.(\ref{relation1}), $\frac{\beta}{\lambda}$ must be small if one requires the inflaton to dominate the inflation. The parameter $V_0$ should equate to $\Lambda^4\approx 3\times 10^{-121}M^4_{\rm
Pl}$, where $\Lambda$ is the cosmological constant. In this paper we consider the values of $\lambda$ and $\beta$ in the below ranges,
\begin{equation}
0.5\leq\lambda\leq 2, \quad\quad\quad   0\leq\beta\leq 0.1.
\end{equation}
Now the value of $C$ is known, and then the value of $A_0$ can be determined in terms of $\lambda$ and $\beta$. The result is in fig.(~1).
\begin{figure}[h]\label{fig:A0}
\centering
\includegraphics[width=7.5cm]{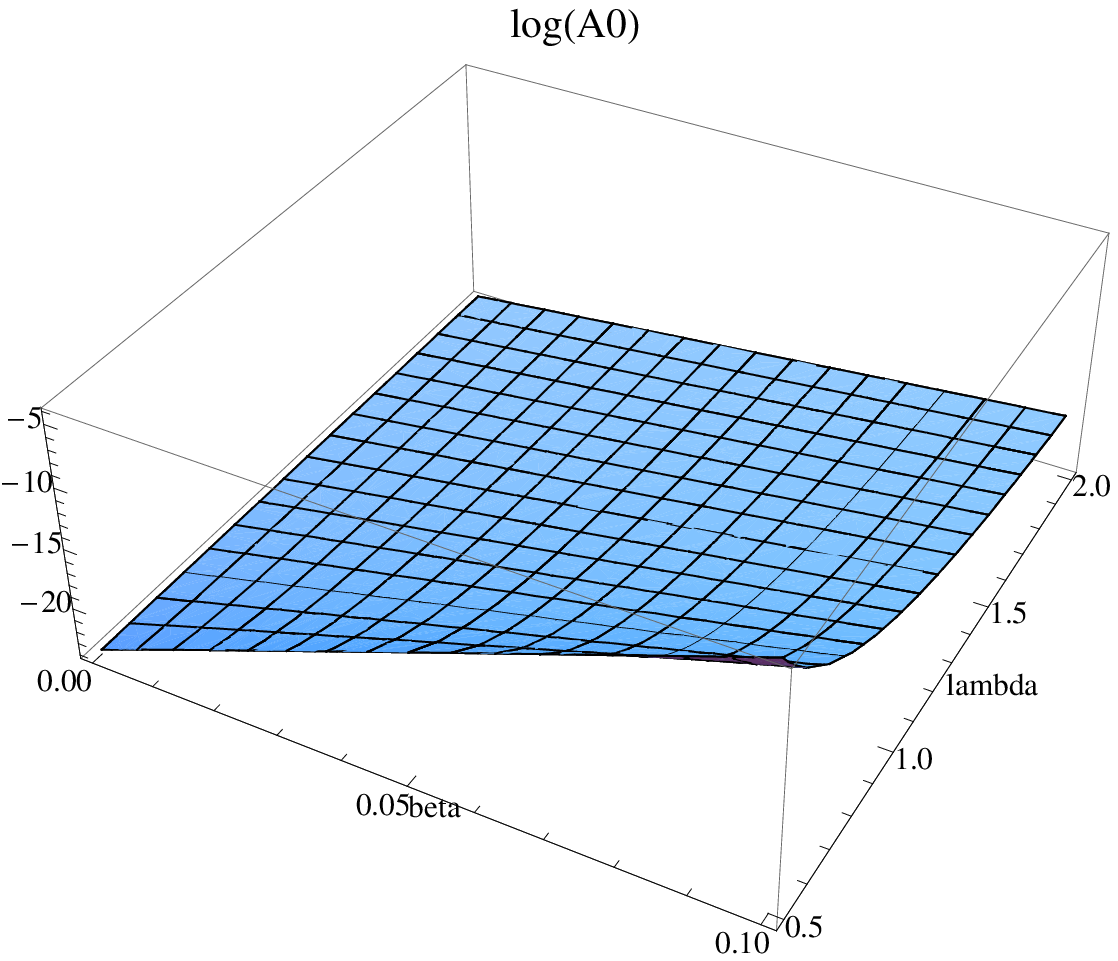}
  \caption{$log(A_0)$}
\end{figure}

Now we can calculate the below parameters,
\begin{equation}
\epsilon\approx 8.32\times10^{-3},  \quad\quad c_s\approx2.63\times10^{-2},\quad\quad \eta\approx7.62\times10^{-3},\quad\quad \kappa\approx3.81\times10^{-3}.
\end{equation}
Using the above results, we can evaluate the power spectrum and spectral index.
\begin{equation}
n_S\approx0.972,   \quad  {\cal P}_{g}\approx 8.45\times10^{-12}, \quad r\approx0.004,\quad  n_T \approx -0.016.
\end{equation}
These results are consistent with the current observed data.
At the end of inflation, the inflaton oscillates around its minimum and decays. After that, the quintessence field recovers its freedom and behaves like a standard canonical field. 

\section{Numerical simulation} \label{sec:5}

In section.(\ref{sec:2}), we considered the theory of k-inflation
coupled with a dark energy field. In the analysis we
assumed slow roll for both fields and neglected the kinetic term of
the Q field in Friedmann equations. We argued the Q field,
despite of different initial values, would quickly settle into the
minimum of the effective potential within a few e-foldings. In this section, we
reserve all of the terms in the equations of motion for the Q field
and the inflaton field as well as in Friedmann equations, and do a
numerical analysis.

Fig.(~2) shows the Q field with different initial values
(marked by blue and green dashed lines) typically evolves toward the minimum of the effective potential
(marked by solid yellow line as calculated by eqn.(\ref{Qmini})). As
inflation continues, the denominator of eqn.(\ref{Qmini}) gradually
declines and the value of Q field in the minimum of effective
potential slowly grows. The numerical simulation is consistent with this
simple analysis-the Q field is attracted to the solid yellow line
and subsequently increases as a function of $N=\ln a$.

\begin{figure}[h]\label{fig:Q1}
\centering
\includegraphics[width=8.5cm]{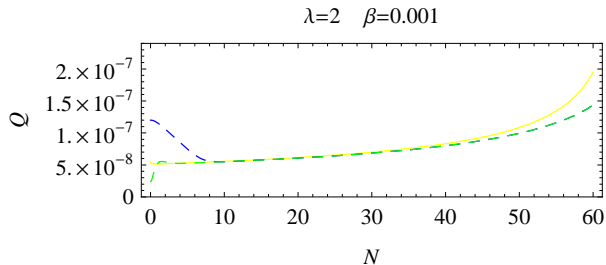}
  \caption{The evolution of $Q$ }
\end{figure}

\section{conclusions} \label{sec:6}

In this paper we described a model of k-inflation coupled with a
quintessence field, which played the role of dark energy. We
chose an inverse power law potential: $V(Q) = V_0 M_{\rm Pl}^4(\frac{Q}{M_{\rm Pl}})^{-\lambda}$ for the dark energy field, and a particular coupling function
$A(Q)=A_0 (\frac {Q}{M_{\rm Pl}})^{\beta}$. We assumed for proper parameters the
dark energy field, despite of varied initial values, would quickly
settle in the minimum of its effective potential. Further we did a strict numerical simulation to justify our assumption about the behavior of the Q field. The effective mass
of the dark energy field is usually heavy
compared with the Hubble parameter during inflation. We studied the perturbations of the inflaton field and found that the speed of sound was modified. We also found the modification to the tensor to scalar
ratio only came from the modification of speed of sound and the slow roll parameter, and was
independent of the coupling function $A(Q)$. 

Due to the particular choice of the quintessence potential
and the coupling function, we find that in a sense our model
suffers from a fine-tuning problem. Technically we note that in
this type of directly-coupled model, in order to treat the models
analytically, we have to choose both of the quintessence potential
and the coupling function  to be of a power law or exponential
form at the same time. The latter case has been investigated in
\cite{slow roll inflation with dark energy:2008}, where both of
the quintessence and inflaton are canonical field. In this model,
the quintessence potential and the coupling function are of the
forms $exp(-\lambda\frac {Q}{M_{\rm Pl}})$ and $exp(-\beta\frac
{Q}{M_{\rm Pl}})$ respectively. Similarly, the amplitude of
curvature perturbation is modified but the situation is better
since the modification is an exponential factor which is always
greater than unit, which in our case, due to the smallness of
$V_0$, one has to fine tune $A_0$ such that the amplitude of the
perturbation would not be suppressed too much. Nevertheless, since
the inverse power law potential is one of the most popular models
for quintessence, it makes our model an interesting complement to
the works in \cite{slow roll inflation with dark energy:2008}.
Obviously, it would be interesting to study the K-inflation with
the quintessence potential and the coupling function being of
exponential forms.

In our case, we chose the K-inflation model among many other kinds
of inflation models. We note that even with other kinds of
inflation models, the fine-tuning problem could not be bypassed in
a simple way. This is because that in this class of  directly
coupled models between the inflaton and the quintessence field,
the modification of the cosmological perturbation is almost
certain to be proportional to $A_0$ and a power of $V_0$. One
lesson from our study is that  the coupling function may affect
the amplitude of the perturbation significantly and therefore is
restricted very much by the experimental data.

\section*{Acknowledgments}
The work was partially supported by NSFC Grant No.10535060,
10775002, NKBRPC (No. 2006CB805905) and RFDP.


\begin{thebibliography}{99}
\bibitem{linde book:1990}
Andrei Linde, Partical Physics and Inflationary Cosmology(Harwood,
Chur, Switzerland,1990) [arxiv:hep-th/0503203]. Andrei Linde,
Inflation and Quantum Cosmology(Academic Press, Boston,1990).

\bibitem{Mukhanov book:2005}
V.~F.~Mukhanov, Physical Foundations of Cosmology(Cambridge
University Press, Cambridge, 2005).

\bibitem{linde lecture:2006}
Andrei Linde, Inflationary Cosmology. Lect.\ Notes Phys.\ {\bf 738},
1-54 (2008).

\bibitem{brandenberger inflation progress and problems:1999}
R.~H.~Brandenberger, Inflationary Cosmology: Progress and Problems.
[arxiv:hep-ph/9910410].

\bibitem{k-inflation:1999}
C.~Armendariz-Picon, T.~Damour, V.~Mukhanov. k-Inflation. Phys.\
Lett.\ B {\bf 458}, 209 (1999) [arXiv:hep-th/9904075].

\bibitem{perturbations in k inflation:1999}
Jaume Garriga, V.~F.~Mukhanov, Perturbations in k-inflation. Phys.\
Lett.\ B {\bf 458}, 219 (1999) [arxiv:hep-th/9904176].

\bibitem{Self-reproduction in k-inflation:2006}
Ferdinand Helmer, Sergei Winitzki, Self-reproduction in k-inflation. Phys.\ Rev.\ D {\bf 74}, 063528 (2006) [arxiv:gr-qc/0608019].

\bibitem{SN}
  A.~G.~Riess et al.[Supernova Search Team Collaboration],
  Astrophys.\ J.\ {\bf 607}, 665 (2004) [arXiv:astro-ph/0402512].
\bibitem{LSS}
  M.~Tegmark et al.,
  Phys.\ Rev.\ D {\bf 69}, 103501 (2004) [arXiv:astro-ph/0310723];
  K.~Abazajian et al.,
  Astrophys.\ J.\ {\bf 129}, 1755 (2005) [arXiv:astro-ph/0410239];
  M.~Tegmark et al.,
  Astrophys.\ J.\ {\bf 606}, 720 (2004).
\bibitem{CMB}
  D.~N.~Spergel et al.,
  Astrophys.\ J.\ Suppl.\ {\bf 148}, 175 (2003) [arXiv:astro-ph/0302209].



\bibitem{dynamics of dark energy:2006}
Edmund J.~Copeland, M.~Sami, Shinji Tsujikawa, Dynamics of dark
energy. Int.\ J.\ Mod.\ Phys.\ D {\bf 15}, 1753 (2006)
[arxiv:hep-th/0603057].


\bibitem{quintessence}
  P.~J.~E.~Peebles and B.~Ratra,
  Astrophys.\ J.\  {\bf 325}, L17 (1988);
  B.~Ratra and P.~J.~E.~Peebles,
  Phys.\ Rev.\ D {\bf 37}, 3406 (1988);
  C.~Wetterich, Nucl.\ Phys.\ B {\bf 302}, 668 (1988);


\bibitem{initial conditions:2002}
Michael Malquarti, Andrew R.~Liddle, Initial conditions for
quintessence after inflation. Phys.\ Rev.\ D {\bf 66}, 023524 (2002)
.

\bibitem{large scale cmb and dark energy}
J.~Weller and A.~M.~Lewis. Large Scale Cosmic Microwave Background
Anisotropies and Dark Energy. Mon.\ Not.\ Roy.\ Astron.\ Soc.\ {\bf
346}, 987 (2003) [arxiv:astro-ph/0307104].


\bibitem{slow roll inflation with dark energy:2008}
Philippe Brax, Carsten van de Bruck, Lisa M.~H.~Hall, and Joel
M.~Weller, Slow-Roll Inflation in the Presence of a Dark Energy
Coupling, Phys.\ Rev.\ D {\bf 79}, 103508 (2009) [arxiv:0812.2843].

\bibitem{five years wmap data:2008}
J.~Dunkley, and his collaborators. Five-Year Wilkinson Microwave Anisotropy Probe (WMAP1) Observations:
Likelihoods and Parameters from the WMAP data, Astrophys.\ J.\ Suppl.\, {\bf 180} (2009) [arxiv:astro-ph/0803.0586].


\bibitem{perturbations:1992}
V.~F.~Mukhanov, H.~A.~Feldman, R.~H.~Brandenberger. Theory of
Cosmological Perturbations, Phys.~Rep. {\bf 115}, (1992).

\bibitem{perturbations lectures:2003}
R.~H.~Brandenberger. Lectures on the Theory of Cosmological
Perturbations, Lect.\ NotesPhys.\ {\bf 646},127 (2004)
[arXiv:hep-th/0306071].

\bibitem{Cosmological constraints on general single field inflation}
Nishant Agarwal, Rachel Bean, Cosmological constraints on general, single field inflation, Phys.\ Rev.\ D {\bf 79}, 023503 (2009)
[arXiv:astro-ph/0809.2798].

\bibitem{Exploring Parameter Constraints on Quintessential Dark Energy: the Inverse Power law model}
Mark Yashar, Brandon Bozek, Augusta Abrahamse, Andreas Albrecht, and Michael Barnard, Exploring Parameter Constraints on Quintessential Dark Energy: the Inverse Power Law Model, Phys.\ Rev.\ D {\bf 79} ,103004 (2009) [arXiv:astro-ph/0811.2253].


\bibitem{stochastic quintessence}
Jerome Martin, Marcello Musso, stochastic quintessence, Phys.\ Rev.\ D {\bf 71}, 063514 (2005) [arXiv:astro-ph/0410190].

\end{thebibliography}
\end{document}